# Accessing Plasmonic Hotspots using Nanoparticle-on-Foil Constructs

Rohit Chikkaraddy*[1] and Jeremy J Baumberg*[1]

[1] NanoPhotonics Centre, Cavendish Laboratory, Department of Physics, JJ Thompson Avenue, University of Cambridge, Cambridge, CB3 0HE, United Kingdom

ABSTRACT.

**Metal-insulator-metal (MIM) nanogaps in the canonical nanoparticle-on-mirror geometry (NPoM) provide deep-subwavelength confinement of light with mode volumes smaller than $V/V_\lambda < 10^{-6}$. However, access to these hotspots is limited by the impendence mismatch between the high in-plane $k_\parallel$ of trapped light and free-space plane-waves, making the in- and out-coupling of light difficult. Here, by constructing a nanoparticle-on-foil (NPoF) system with thin metal films, we show the mixing of insulator-metal-insulator (IMI) modes and MIM gap modes results in MIMI modes. This mixing provides multi-channel access to the plasmonic nanocavity through light incident from both sides of the metal film. The red-tuning and near-field strength of MIMI modes for thinner foils is measured experimentally with white-light scattering and surface-enhanced Raman scattering from individual NPoFs. We discuss further the utility of NPoF systems since the geometry allows tightly confined light to be accessed simply through different ports.**

**Graphic for Manuscript**

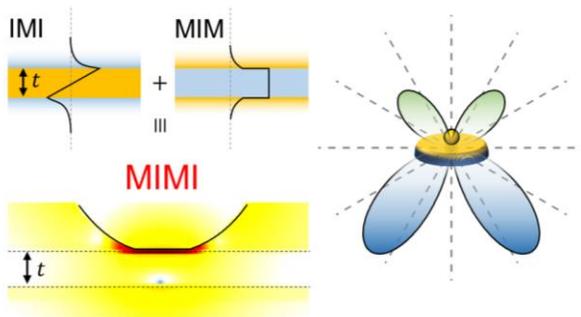

Despite the recent realization that visible light can be trapped using metals to the nanometer scale and below (mode volumes $V < 100$ nm³), efficient access to these 'hotspots' gives many challenges[1,2]. Essentially the requirement is for impedance matching through an antenna[3–5], so that coupling is then much better than the simple overlap integral $V/\lambda^3 < 10^{-6}$. While surface plasmon polaritons (which confine on the 10-100nm length scale) can be accessed using grating or prism couplers, this is not efficient with tighter confinement[6–9] because the in-plane $k_\parallel$ required is so large.

At the heart of metal-confining light architectures are layered structures with subsequent lateral patterning. Typically, these layers are based around two variants: (i) Insulator-metal-insulator (IMI) sheets, or (ii) metal-insulator-metal (MIM) gaps. In IMI systems, SPPs on either side of a metal film strongly hybridize when the metal thickness ($t$) is smaller than its skin depth (<20nm), resulting in large in-plane wave-vector ($k_\parallel$) modes confining EM fields close to the metal sheet. Recently such modes were identified in 2-dimensional graphene and metallic transition metal dichalcogenide (TMD) monolayers above their plasma wavelength (4-12µm), and were utilized to achieve high confinement as well as molecular sensing[10–13]. In MIM gaps, the SPPs on either side of the dielectric medium similarly hybridize to form symmetric gap modes with large $k_\parallel$. Such confined light between two metal surfaces in a narrow gap has been extensively utilized for sensing[14,15], Purcell-enhanced emission from single emitters[16,17], strong coupling[18], surface enhanced Raman scattering (SERS)[19] and in metamaterials[20].

Crucial to obtain tight confinement in both cases is to achieve reliable sub-nm film thicknesses ($t$) or gaps ($d$) capable of producing large $n_{\text{eff}}$ (>100) which corresponds to effective plasmon wavelengths of <10nm. While the thinnest metal sheets have reached $t \sim 3$nm,[21–23] it has been possible using TMD or molecular spacers to routinely achieve $d<1$nm gaps[24]. Convenient lateral confinement at the 10nm scale can then be produced using nanoparticle (NP) facets created using colloidal assembly (in MIMs between NPs in dimers[25–27]), though top-down fabrication also aims to access the same domain[28,29]. A particularly reliable, scalable and robust construct for MIM geometries uses nanoparticle-on-mirror (NPoM) systems[30,31] where a colloidally-synthesized (usually single-crystal gold) nanoparticle is placed on top of an atomically flat >100nm-thick mirror which is pre-deposited with a self-assembled monolayer of molecules. This results in an MIM geometry with gap size set by the thickness of the molecular layer ($d<2$nm), identical across large areas (>100cm²)[32]. However, access to these hotspots is confounded by the thick non-transparent Au mirror, restricting its utilization.

In this work, we present a nanoparticle-on-foil geometry (NPoF) with a finite thickness Au-substrate providing both front and back access to the confined light in the nano-gap. It arises from the mixed coupling of both IMI and MIM modes across the thin Au foil, resulting in new metal-insulator-metal-insulator (MIMI) modes. By tuning the film thickness, we modify the effective index of these MIMI gaps and control the far-field scattering and near-field SERS.

In previous work on related NPoF geometries with film thickness >20nm or gaps >5nm, the nanocavity modes were not understood to depend on film thickness even though the gap is accessible with the excitation of surface plasmon polaritons (SPPs) on the metal film through prism coupling[33–36]. By making even thinner metal films, we might expect the effective image dipole coupling would fade out and coupled plasmon cavity modes would damp out. However surprisingly, we find in NPoF systems here with $t<10$nm and $d<2$nm that the coupled SPPs on either side of the foil mix with MIM modes trapped in the gap forming MIMI modes.

**Coupling of IMI and MIM modes**

Before discussing the coupling of MIM and IMI modes (Fig.1a,b), we need to compare the key differences between MIM and IMI plasmon mode dispersions. The explicit IMI dispersion in the thin film limit is given by[37,38]:

$$k_\parallel = k_0 \sqrt{\varepsilon_d + \frac{4}{\varepsilon_d \varepsilon_m (k_0 t)^2}} \quad \text{(Eq. 1)}$$

where $k_o = \frac{2\pi}{\lambda}$. Here $\varepsilon_m(\omega)$ and $\varepsilon_d$ are the dielectric functions of the metal and the dielectric respectively. The values obtained from the above equation match well with the full solution (SI). The dispersion of an MIM mode in the narrow gap limit is given by[39–41] (see Supporting Information A)

$$k_\parallel = k_0 \sqrt{\varepsilon_d + 2\zeta \left(1 + \sqrt{1 + \frac{\varepsilon_d - \varepsilon_m}{\zeta}}\right)}; \quad \zeta = (k_0 d \varepsilon_m / \varepsilon_d)^{-2} \quad \text{(Eq. 2)}$$

When $d$ or $t$ is large (>50nm), these dispersions are not significantly different from that of an SPP on a single Au surface. However, when $d, t$ <5nm, the dispersions become flat with extremely small plasmon wavelengths (< 10nm). For $d = t$, IMI and MIM dispersions converge to the same $n_\text{eff} = k_\parallel/k_0$ at large $k$ (Fig.1c).

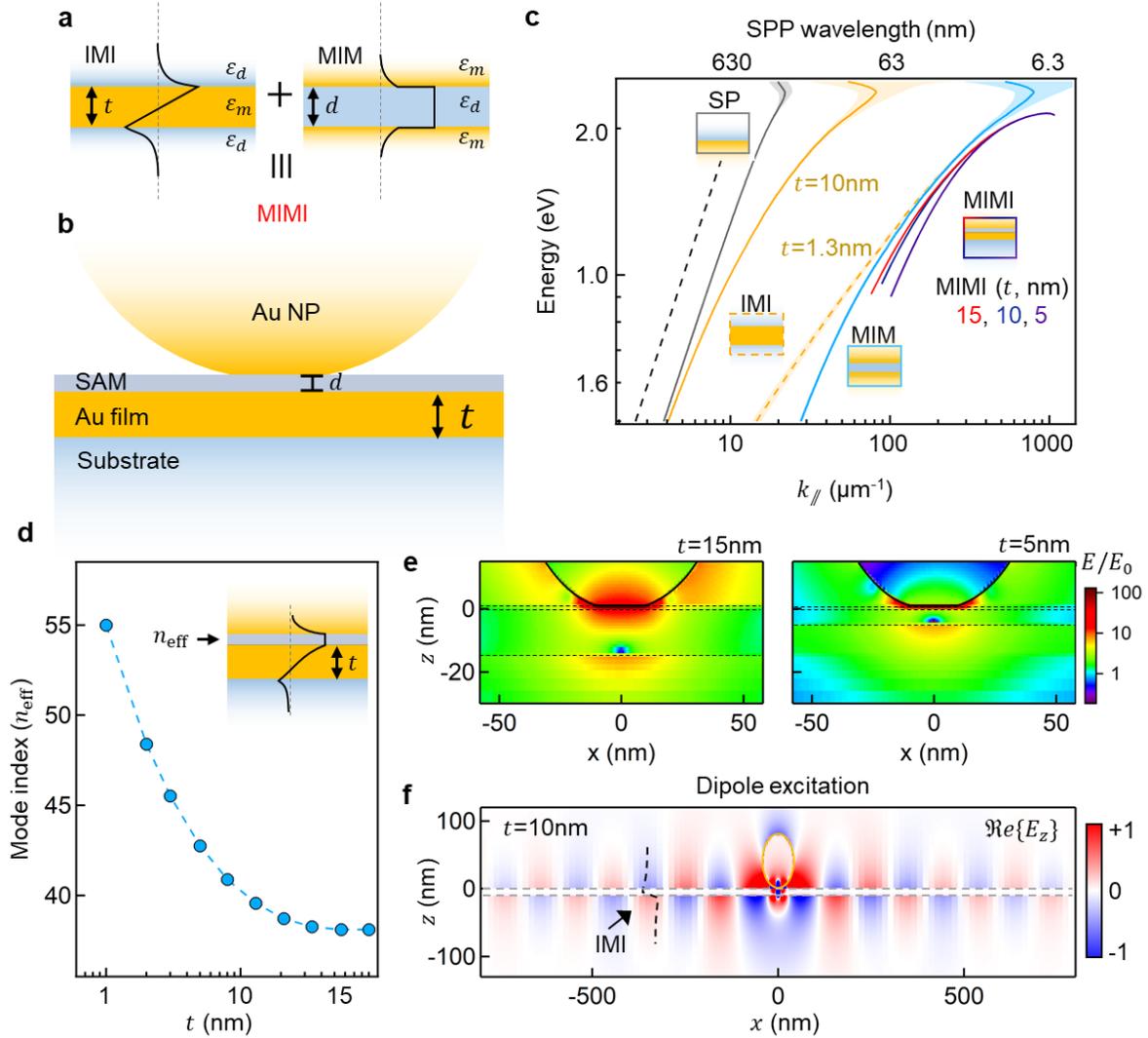

**Figure 1.** Coupling of IMI + MIM = MIMI modes in a nanoparticle-on-foil cavity. (**a**) Tight mode confinement in IMI and MIM geometries. Black curves show $H_y$ in thin film ($t$) and narrow gap ($d$) limits. (**b**) Schematic nanoparticle-on-foil (NPoF) geometry here combines MIM and IMI modes into a single architecture of hybridized MIMI modes. (**c**) Dispersion of MIMI (red-purple) modes for different $t$ with $d$=1nm compared to IMI (yellow, $t$=10nm solid, $t$=1.3nm dashed) and MIM (blue, $d$=1.3nm) modes; surface plasmon on thick Au (grey) and free-space photon (dashed grey). (**d**) Effective propagation index of MIMI modes for increasing $t$ at $\lambda$=633nm. (**e**) $E_z$ near-fields from full-wave simulations of NPoF with plane wave excitation for $t$=15nm and $t$=5nm film thickness at the nanocavity resonance. The field null in the metal-film just above the substrate and the leakage of nanocavity fields into the substrate are clearly seen. (**f**) $E_z$ near-fields from full-wave simulations of NPoF geometry with dipole source exciting the NPoF gap centre for $t$=10nm at $\lambda$=633nm, highlighting IMI modes travelling away from the AuNP.

In the NPoM geometry, the facet width $w$ of the AuNP on top defines the lateral discretization of the MIM modes. Essentially, this forms a type of Fabry-Perot resonator with solutions $k_\parallel w = n\pi$ for 1D (and similar Bessel function solutions in 2D)[42,43] formed by the reflections on either end due to the large mismatch between $k_\parallel$ in the MIM gap and $k_0$ in free space[42,44]. However, these resonant hotspots in the gap under the NP are relatively inaccessible and typically emit into high angles for small $d$, thus compromising efficient coupling.

This can be modified through hybridizing MIM and IMI modes by reducing the metal thickness to $t$<20nm (Fig.1c). The IMI fields at the dielectric substrate/Au interface then interact with the MIM gap fields trapped in the AuNP-gap-Au film resulting in MIMI modes with $n_{\text{eff}}$ >50 for $t$<5nm (Fig.1d). Full wave simulations for NPoFs with a dipole excitation source in the middle of the MIM gap confirms the launching of IMI modes away from the NPoF (Fig.1e,f). The effective wavelength of IMI modes is tuned as the film thickness reduces (Supporting Information, Fig.S1). It is important to note that without the AuNP, the magnitude of IMI fields launched is 8000x lower and this is because the radiation from the dipole in the MIM gap is Purcell enhanced.

As the gap reduces, the intense local field inside the MIM gap leaks further underneath the foil producing an accessible hotspot in the dielectric substrate (Fig.1e). As we show below, this can be made available for a variety of coupling schemes.

**Far-field scattering from MIMI gaps**

To characterize optical coupling into these localized MIMI modes we fabricate NPoF samples with different thicknesses of the Au foil on a glass substrate. The AuNPs of diameter $2R$=80nm are deposited on top of a compact self-assembled monolayer of BPT molecules ($d$=1.3nm) and the dark-field scattering spectra from individual Au NPoFs are collected using a custom-built white light microscope. For the $t$=20±2nm foil, the dominant scattering resonance is observed at 805±10nm but when the foil thickness is reduced to 10nm, the resonance shifts to 890nm (Fig.2a,b). Intuitively this redshift is surprising because a thinner confining metal foil under the NPoM would be expected to decrease the optical confinement, thus leading to blue shifts. The origin of this is the asymmetric MIMI mode shape (Fig.1d inset) which, in fact increases the confinement due to the field null inside the metal foil. Intuitively this can be also understood from the mixing between IMI and MIM modes that produces a lower energy coupled mode, thus shifting it to higher wavenumber (Fig.1c) that results in a faster exponential decay into the metal thus increasing confinement[30]. The redshift is >10 standard deviations outside the spectral shifts arising from AuNP size and shape variations, while the observed scattering intensity is 10× lower for $t$=10nm foils compared to $t$=20nm at resonance.

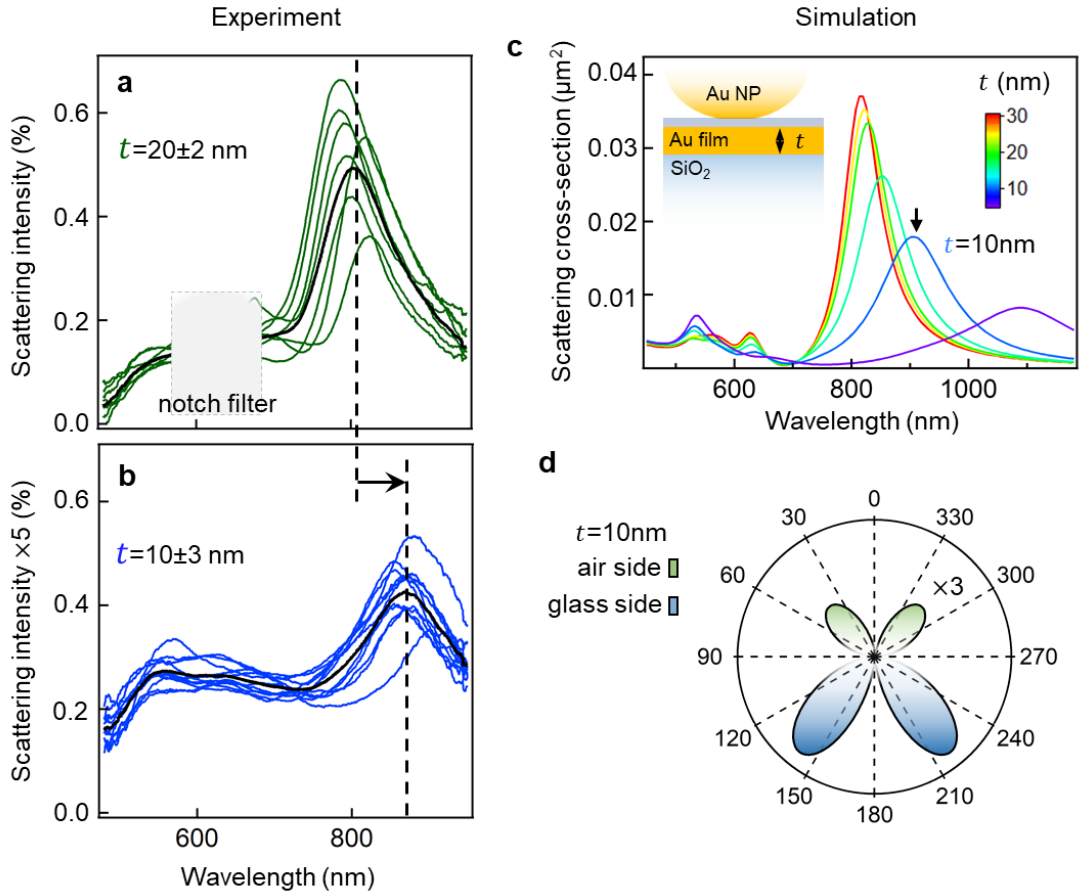

**Figure 2.** Far-field scattering from NPoFs. (**a,b**) Dark-field scattering spectra obtained from 10 different individual NPoF cavities for (a) $t$=20nm and (b) $t$=10nm. The black curve represents the average spectra. The redshift in the scattering resonance is indicated by the horizontal arrow; notch laser filter for Raman shown in grey. (**c**) Simulated scattering spectra for NPoF cavities *vs* decreasing foil thickness $t$ from 30nm to 5nm (black arrow marks $t$=10nm). (**d**) Simulated far-field radiated intensity on a polar plot for NPoF cavity, showing both air-side and glass-side emission.

Two key features are observed in the experimental data shown in Fig. 2a,b. Firstly, the shift in the plasmon resonance with foil thickness and secondly, the change in the mode intensity. These experimental observations are corroborated with 3D full-wave simulations (see Methods) using the experimental geometry. As $t$ is tuned from 30nm to 10nm the NPoF scattering resonance shifts gradually from 815nm to 905nm (Fig. 2c), matching the experiments. A further reduction in $t$ shifts the resonance beyond $\lambda$=1µm. This redshift is consistent with the estimated change in $n_\text{eff}$ for MIMI modes, and can be understood from the MIM to MIMI dispersions (Fig.1d) which for fixed $k_\parallel = \pi/w$ yield redshifted resonances. The ±10nm discrepancy observed between simulation and experiment is associated with the exact nanoparticle shape and gap morphology. A poorer agreement is found for scattering intensities, with simulations showing only a two-fold decrease on halving $t$ which contrasts with 10-fold reduction observed in experiments. To understand this, we calculate the far-field radiation pattern of light scattering from an NPoF to both air- and glass-sides (Fig.2d). Unsurprisingly, the light now leaks into the glass side beyond the total-internal reflection angle (TIR=45°) when $t$ is below the skin depth of Au (10nm). For $t$=10nm the light scattered into the air is 4.5× times smaller than the light leaking to the glass-side, which thus accounts for the observed 10-fold decrease in experiments (scattered light is collected only on the air-side with NA=0.9). We also confirm this experimentally using a modified dual channel microscope with white laser illuminating wide-field areas of the sample at high incident angles and the scattered light collected separately on air and glass sides (Supporting Information, Fig. S2). Note that the fraction of light leaking to the glass-side can reach >90% for $t$<5nm.

**Near-field and SERS**

To characterize the near-field strength of these modes, surface-enhanced Raman scattering (SERS) signals are obtained from the self-assembled monolayer of biphenyl-4-thiol (BPT) molecules assembled between the AuNP and Au-film. Individual NPoFs are illuminated from the air-side with a 633 nm laser and backscattered Stokes light is measured.

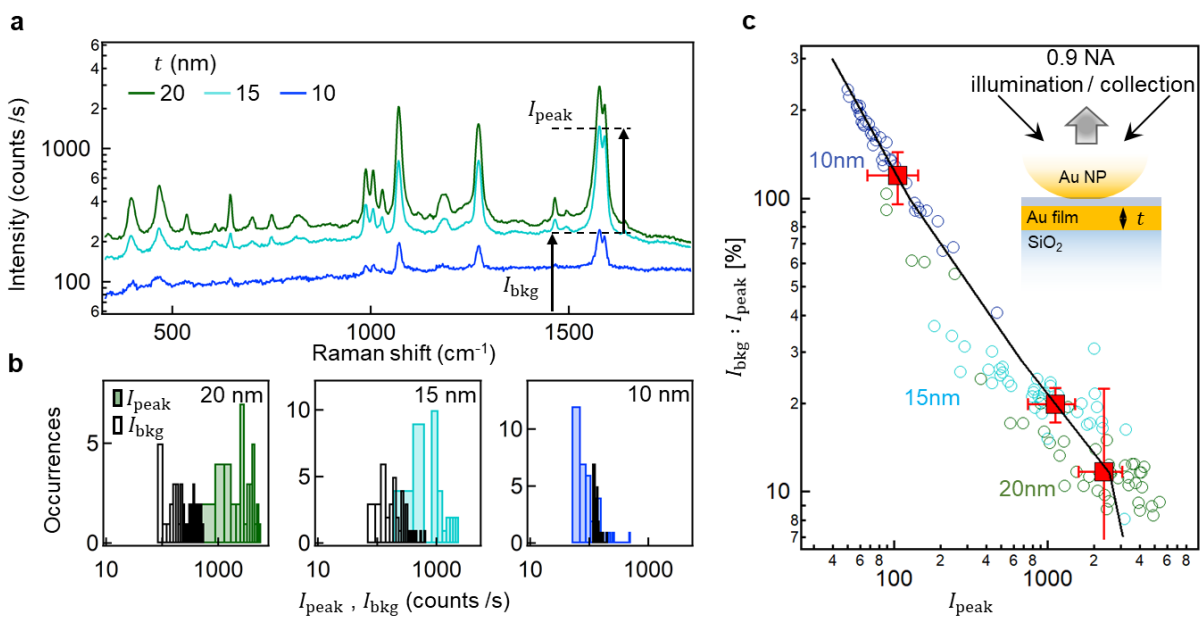

**Figure 3.** SERS enhancements in NPoFs. (**a**) Average SERS spectra obtained from 30 individual NPoF for 3 different film thicknesses. The SERS intensity is plotted on a log-scale. (**b**) Histogram of extracted SERS 1585 cm$^{-1}$ peak and background intensities for 30 individual NPoFs. (**c**) Variation in the ratio of SERS background to SERS peak intensities. Data from different $t$ are color coded to (b) and average values for each $t$ are shown as red points along with their standard error. Black line shows prediction from full wave-simulations.

The average SERS spectra from 30 different NPoFs for 3 different foil $t$ are shown in Fig. 3a (note SERS intensity is plotted on a log-scale). The observed SERS signals contain sharp narrow peaks of widths <15cm$^{-1}$ corresponding to different Raman vibrational modes of BPT molecules and a broad inelastic background from the electronic Raman scattering (ERS) of conduction band electrons in both AuNP and Au foil. Intensities from these two processes are quantified by Gaussian fits to the BPT vibrations to extract their peak height ($I_{peak}$) as well as the SERS background ($I_{bkg}$). Histograms of $I_{peak}$ and $I_{bkg}$ from 30 NPoFs reveal a key feature that as $t$ decreases from 20nm to 10nm, the overall SERS intensity dramatically decreases and $I_{peak}$ drops by >20×. This cannot be attributed to the shift in plasmon resonance position, as on- *vs* off-resonance excitation changes the intensity only by a factor of 5.

For practical sensing the important quantity is the ratio of peak SERS to background, as well as how this ratio responds to the tuning of fields inside and outside the metal. To quantify these variations, the ratio $I_{bkg}:I_{peak}$ is plotted *vs* $I_{peak}$ (Fig. 3c). While for $t$=20nm there is only 10% contribution from the metal electronic SERS compared to the molecular SERS, this reaches $I_{bkg} \geq I_{peak}$ for $t$=10nm, arising from the greater penetration of light into the Au foil, as shown below.

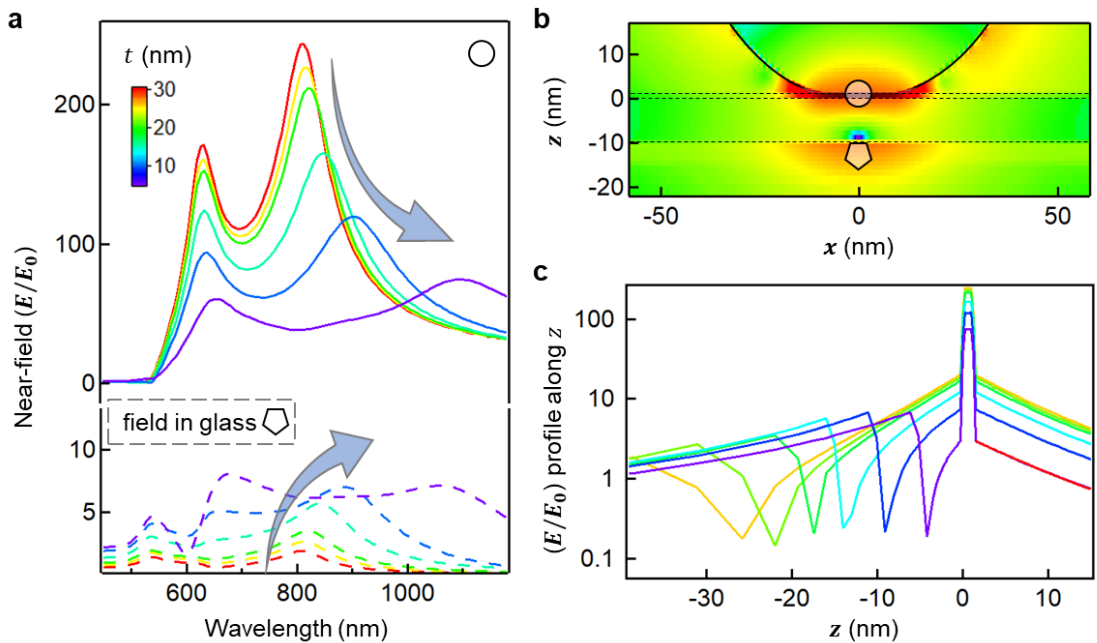

**Figure 4.** Near-field enhancements in NPoFs. **(a)** Simulated wavelength-dependent near-field intensities extracted from the centre of the BPT-gap layer and fields 2nm below the Au-foil (indicated by circle and pentagon symbols in (b)) for different film thickness (NPoF is illuminated with plane wave with polarization perpendicular to the film surface at a glancing angle 90°). **(b)** Near-field enhancement map for $t$=10nm at the nanocavity resonance. **(c)** Extracted near-field profile along $z$-axis (at $x$=0) for different film thickness, clearly highlighting the fast decay of fields in the metal-film.

Near-field enhancement factors are extracted from wavelength-dependent full-wave simulations (see Methods). The near-field enhancement ($E/E_0$) extracted from the centre of the NPoF with a BPT monolayer decreases for smaller $t$ and follows closely the scattering cross section (Fig. 2c). The simulations for normal illumination show similar tends (Supporting Information, Fig. S3). The change in $I_{\text{SERS}} \propto (E_{in}^2/E_0^2) \cdot (E_{out}^2/E_0^2)$ in the in/out-going wavelength range 600-750nm multiplied by the fraction directed to the air-side accounts for the 25-fold decrease in the experimental SERS intensity as $t$ is decreased from 20nm to 10nm (Fig. 3c, Supporting Information, Fig. S4). It is also apparent that the NPoF cavity resonance is slightly more damped for thin foils, but still clearly present.

This near-field also allows the scaling between $I_{\text{bkg}}$ and $I_{\text{SERS}}$ to be derived. The decay length ($\delta$) of light confined inside the Au on either side of the gap is obtained from an exponential fit to $(E/E_0)^2$ for the fields inside the AuNP (Supporting Information, Fig. S4). Since the field penetrating the metal then scales as $E/t$, the ratio of bgd:SERS $\propto (Et^{-1}/E)^4 \cdot (t/d) \propto t^{-3}$ which matches well the experimental data (Fig. 3c), with full simulations giving the line shown. Our measurements thus corroborate the MIMI mixing model, and indicates that while the field in the gap halves when the foil drops to 10nm, the field in the dielectric underneath increases tenfold (at 633nm, Fig.4). The field null inside the foil is a direct consequence of IMI and MIM mode mixing (Fig.4c). As we derive, the dispersion of coupled MIMI modes shows an in-plane momentum ($k_\parallel$) which is larger compared to MIM and IMI modes (Fig.1c), which is what leads to stronger field confinement. This can be intuitively understood from the anticrossing of MIM and IMI modes, which pushes one coupled mode to lower energy (and thus indeed to high wavevectors). Since $k_\perp^2 = k_0^2 - k_\parallel^2$ and thus $k_\perp \sim ik_\parallel$, modes with larger wavevector decay faster into the metal, increasing the light confinement. As the foil thickness decreases, the hotspot thus becomes more accessible with improved radiative coupling, but the penetration of light into the metal and the greater field gradient enhances ERS metal scattering. In some experiments, this can cause problems (for instance S:N values for molecular sensing) while in others it can prove beneficial, for instance in nonlinear optical mixing from plasmons.

**Conclusion and outlook**

Having both front and back access to these NPoF systems facilitates improved active control of the material properties assembled into the plasmon proximity. To highlight this, we show here a variety of examples.

One long-sought opportunity has been to use magnetic control of the gap material. The NPoF can be assembled onto a magnetic substrate such as iron (Fe) without damping the plasmons when $t$=10-15nm (Fig. 5a, Supporting Information, Fig.S5). This allows a wide variety of magneto-plasmonic applications in controlling molecular and semiconductor spin systems, for example, in transition metal dichalcogenides (TMDs) and other monolayers and bilayers.

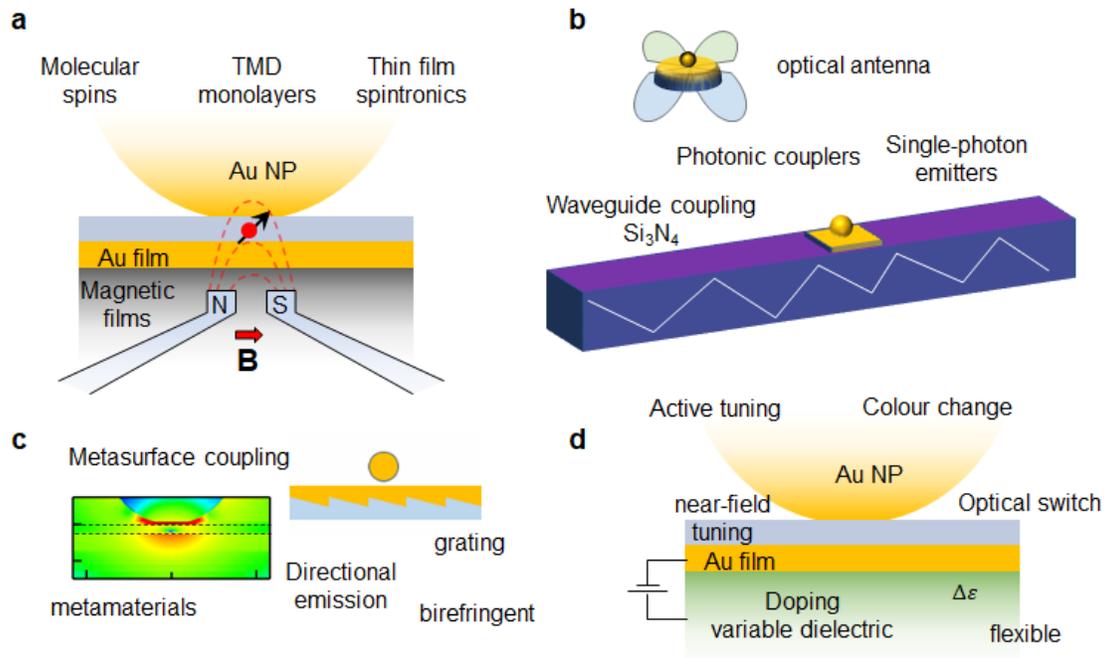

**Figure 5.** Utilization of NPoF geometry. (**a**) Coupling of magnetic and plasmonic spin systems. (**b**) Waveguide integration of NPoFs for single-emitters, SERS sensing, and quantum information. (**c**) Photonic modification of thin metal surface allows coupling of a metasurface to a nanocavity. (d) Doping of underlying dielectric medium changes the MIMI modes and tunes the color as MIMI mixing occurs for $d \sim 10nm$ gaps.

Another advantage is providing back-access to radiation from NPoFs, allowing their integration into waveguides which provide coupling to single emitters inside the NPoF gaps (Fig. 5b). The combination of the ultra-narrow gap geometry providing high Purcell enhancements and the effective coupling to dielectric waveguides, make the system highly viable for quantum optics and sensing.

Importantly, the NPoF construct need not be limited to flat thin Au surfaces. The back-surface of Au can be modified further with photonic/plasmonic metamaterial architectures to fine tune the IMI modes to create other types of MIMI modes (Fig.5c). This parameter space allows the tuning of the scattered radiation rates and directions, and enables new designs incorporating flat lenses. The tunability of this construct makes it compatible to fabricate on flexible surfaces, thus allowing large-scale manufacturing on roll-to-roll systems. As a further opportunity, the dielectric medium underneath the Au surface can also be voltage-tuned (Fig.5d). An example of such an actively-changing dielectric would be to integrate 2D materials such as graphene to make this function as a low energy switch.

The nanoparticle-on-foil geometry thus supports new MIMI modes and provides multi-channel accessibility to plasmonic hotspots which gives advantageous opportunities for enhanced integration of extreme plasmonics into nanophotonics devices.

**METHODS:**

**Sample preparation.** Au is deposited directly on a clean $SiO_2$ microscope coverslip with a deposition rate of 1 Å/s (LEV Lesker, e-beam evaporator) of different thickness after coating with 2nm of Cr as an adhesive layer. For the realisation of single monolayers (SAMs), the Au-coated sample pieces are dipped in 1 mM solution of biphenyl-4-thiol (BPT, Sigma Aldrich, 97%) in anhydrous ethanol (Sigma Aldrich, <0.003% $H_2O$) for 12 h. Nanoparticles of 80nm in diameter (BBI Solutions) are deposited directly onto the BPT treated Au surface. The deposition time is 15 s.

**Experimental setup.** The sample is placed on a motorised stage (Prior Scientific H101) which is fully automated using an in-house code written in Python. We used an Olympus BX51 microscope with a long working distance ×100 NA 0.8 objective. A spectrally filtered 632.8 nm diode laser (Matchbox, Integrated Optics) with 100μm/μm$^2$ power on the sample and spectral linewidth of 0.1 pm is used as the excitation pump. In SERS experiments, we filter laser light with a pair of notch filters centred at 633 ± 2nm (Thorlabs). Inelastically scattered light from the nanoconstructs is coupled through a tube lens into an Andor Shamrock i303 spectrograph and a Newton EMCCD. For dark-field measurements, we used a halogen lamp to excite our samples. Note that we keep the lamp on for around 30 min to stabilise the lamp's power before starting measurements. The reflected light is collected through the same objective and split to an imaging camera (Lumenera Infinity3-1) and a fibre-coupled spectrometer (Ocean Optics QEPRO) for dark-field spectroscopy.

**Numerical simulations.** Full-wave 3D simulations are performed using Lumerical FDTD Solutions v8.12. The Au NP was modeled as a truncated sphere (with facet width of 20nm) of radius 40nm on top of an infinite dielectric sheet of refractive index 1.45 and thickness of 1.3nm matching the BPT thickness. Underneath the BPT layer are different thicknesses of Au film placed above a thick $SiO_2$ substrate. The NPoF is illuminated with plane waves of polarization perpendicular to the film surface at a glancing 90° angle. The fields scattered into the glass and air sides are captured by 2D near-field power monitors.

ASSOCIATED CONTENT

**Supporting Information**. The Supporting Information is available free of charge at http://pubs.acs.org.

(1) Theory of MIM waveguide model and simulation of IMI modes travelling away from the AuNP with different effective index. (2) Experimentally measured light scattering from air and glass slide from individual NPoF cavities in modified dual channel microscope. (3) Normal illumination optical characterization of NPoF. (4) Estimating optical field strength inside the metal for MIMI modes. (5) The light scattering from NPoF onto a magnetic substrate.

AUTHOR INFORMATION

**Corresponding Author**

* Dr Rohit Chikkaraddy, rc621@cam.ac.uk; http://orcid.org/0000-0002-3840-4188

* Prof Jeremy J Baumberg, jjb12@cam.ac.uk; http://orcid.org/0000-0002-9606-9488

**Notes**

The authors declare no competing financial interest.

ACKNOWLEDGMENT

We acknowledge support from the European Research Council (ERC) under the Horizon 2020 Research and Innovation Programme THOR (829067) and POSEIDON (861950) and PICOFORCE (883703). We acknowledge funding from the EPSRC (Cambridge NanoDTC EP/L015978/1, EP/L027151/1, EP/S022953/1, EP/P029426/1, and EP/R020965/1). R.C. acknowledges support from Trinity College, University of Cambridge.